\documentclass{aa}
\usepackage{graphicx}
\begin{document}
\title{Lead abundance in the uranium star CS\,31082-001
\thanks{Based on observations obtained with the Very Large Telescope
 of the European Southern Observatory at Paranal, Chile}}
\author{B. Plez\inst{1} 
\and V.Hill\inst{2}
\and  R. Cayrel\inst{3}
\and M. Spite \inst{2}
\and B. Barbuy\inst{4}
\and T.C. Beers\inst{5}
\and P. Bonifacio\inst{6}
\and F. Primas\inst{7}
\and B. Nordstr\"{o}m\inst{8,9}}
\institute{GRAAL, CNRS UMR 5024, Universit\'{e} de Montpellier 2, 
34095 Montpellier Cedex 5, France 
\and Observatoire de Paris, GEPI, CNRS UMR 8111, 5 place Jules Janssen, 
92195 Meudon Cedex, France 
\and Observatoire de Paris, GEPI, CNRS UMR 8111, 61 av. de l'Observatoire, 
75014 Paris, France
\and IAG Universidade de S\~{a}o Paolo, Dep. de Astronomia CP 3386,
 Rua do Mat\~{a}o 1226, S\~{a}o Paolo 05508-900, Brazil 
\and  Department of Physics \& Astronomy and JINA, Michigan State 
University, East Lansing, MI 48824 USA 
\and Istituto Nazionale per l'Astrofisica -- Osservatorio Astronomico
di Trieste via G.B. Tiepolo 11, 34131 Italy
\and ESO, Karl-Schwarzschild-Str. 2, 85749 Garching bei M\"{u}nchen,
Germany   
\and Lund Observatory, Box 43, 22100 Lund, Sweden
\and Niels Bohr Institute, Juliane Maries Vej 30, DK-2100 Copenhagen, Denmark}
\offprints{R. Cayrel, \email{roger.cayrel@obspm.fr}}
\date{Received/Accepted}
\abstract{
In a previous paper we were able to measure the abundance of uranium
and thorium in the very-metal poor halo giant BPS
\object{CS\,31082-001}, but only obtained an upper limit for the
abundance of lead (Pb).  We have got from ESO 17 hours of additional
exposure on this star in order to secure a detection of the minimum
amount of lead expected to be present in \object{CS\,31082-001}, the
amount arising from the decay of the original content of Th and U in
the star. We report here this successful detection.  We find an LTE
abundance $\log\mathrm{(Pb/H)}+12=-0.55 \pm 0.15$ dex, one dex below the
upper limits given by other authors for the similar stars
\object{CS\,22892-052} and \object{BD\,+17\degr\,3248}, also enhanced
in {\it r}-process elements. From the observed present abundances of
Th and U in the star, the expected amount of Pb produced by the decay
of $^{232}$Th, and $^{238}$U alone, over 12 -- 15 Gyr is $-0.73\pm
0.17$ dex. The decay of $^{235}$U is more difficult to estimate, but
is probably slightly below the contribution of $^{238}$U, making the
contribution of the 3 actinides only slightly below, or even equal
to, the measured abundance. 
The contribution from the decay of $^{234}$U has was not included, 
for lack of published data. In this sense our determination is
a lower limit to the contribution of actinides to lead production.
We comment this result, and we note that
if a NLTE analysis, not yet possible, doubles our observed abundance,
the decay of the 3 actinides will still represent 50 per cent of the
total lead, a proportion higher than the values considered so far in
the literature.
}
\titlerunning{Lead abundance in \object{CS\,31082-001}}
\authorrunning{Plez et~al.}
\maketitle

\section{Introduction} 

The detection of uranium in an old, very metal-poor star of the
galactic halo, BPS \object{CS\,31082-001}, was first reported in
Cayrel et al. (\cite{CHB01}). A greatly improved analysis, Hill et
al. (\cite{HPC02}), (quoted as paper I) was made possible by a
redetermination of crucial atomic data by Nilsson et al. (\cite{NIJL02,
NZL02}).  Hill et al. have determined the abundance of
U~($\log(\mathrm{U/H})+12. =-1.92 \pm 0.11$) and of Th
($\log(\mathrm{Th/H}) + 12 = -0.98 \pm 0.05$), in the usual scale
$\log(n_H)=12.0$, in \object{CS\,31082-001}. These abundances have
been used as cosmo-chronometers, comparing them to theoretical
estimates of the initial production ratio. The time $ \Delta t$ in Gyr
elapsed from the formation of the two actinides to now, is linked to
the production ratio $\mathrm{(U/Th)_0}$ and the ratio measured in the
star $\mathrm{(U/Th)_{now}}$ by the simple relation :
$ \Delta t = 21.76[ \log(\mathrm{(U/Th)_0}) - \log(\mathrm{(U/Th)_{now}})]$ 
where the coefficient 21.76 is derived
from the half-lives of $^{232}$Th and $^{238}$U.  The superiority of
the pair U/Th over the pair Th/Eu has been amply demonstrated, for
example in Goriely \& Clerbaux (\cite{GC99}), Goriely \& Arnould
(\cite{GA01}), or Wanajo et al. (\cite{WII02}, see their fig.7).  As
both U and Th decay to the stable element lead, it is of great
interest to know the abundance of lead in the star. In Hill et
al. (\cite{HPC02}), only an upper limit to the lead abundance was
given, and we report here the result of a new observation obtained at
ESO Paranal to get this abundance. The time requested was 17 hours,
enough to detect the minimum amount of lead coming from the decay of
the observed elements $^{238}$U and $^{232}$Th into $^{206}$Pb, and
$^{208}$Pb, respectively. In addition, lead may come from other
channels, in particular from the decay of $^{235}$U into $^{207}$Pb,
and from more unstable nuclides decaying very quickly to lead, such as
$^{234}$U.
  
\section{Observations and reduction procedure}

The observations were carried out with the ESO VLT using the UVES
spectrograph with image slicer \#2, 
leading to a spectral resolution of $\simeq$80\,000. 
A total of 13 exposures were collected in service mode,
reaching a total exposure time of 17 hours. The signal-to-noise ratio
of the combined spectrum is around 600 per pixel. The data were
reduced using the standard UVES pipeline (Ballester et al.
\cite{bal00}).

\begin{figure*}
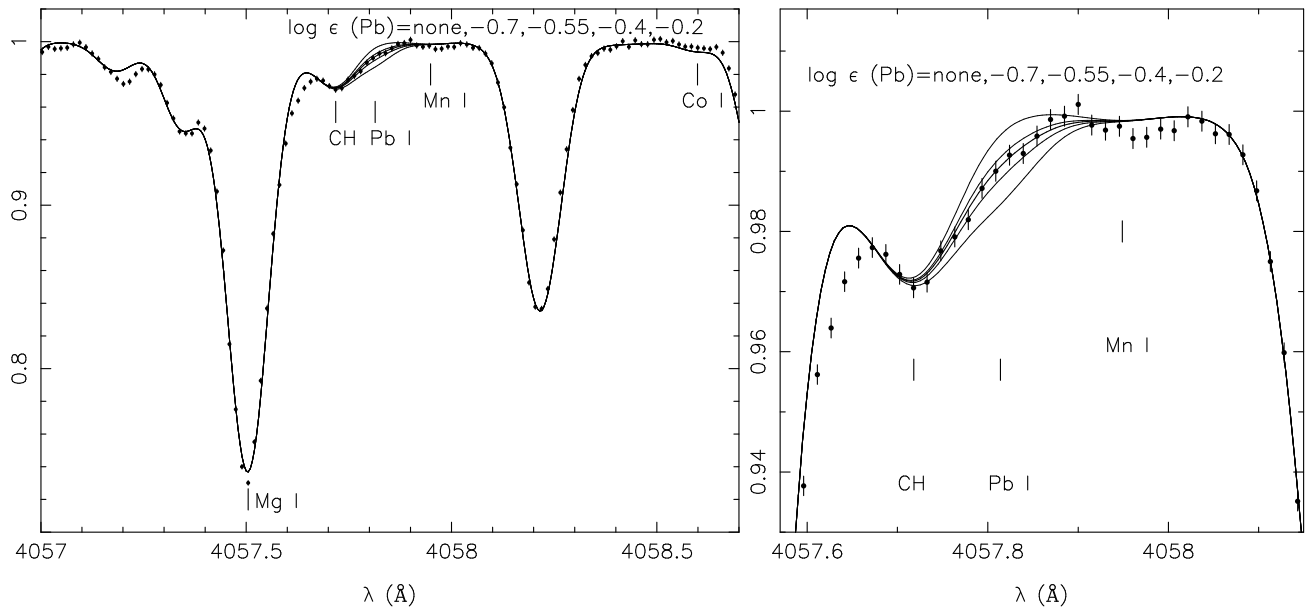

   \centering
  {\includegraphics[angle=0,height=8cm]{Gg211_f1a.ps}
   \includegraphics[angle=0,height=8cm]{Gg211_f1b.ps}}
    \caption{The observed Pb~I 4057.8\ \AA\ line in \object{CS\,31082-001}.
{\it Dots:} Observed (combined) spectrum, with the photon noise
plotted as error bars; {\it lines:} Synthetic spectra computed for the 
abundances indicated in the figure. The {\it left panel} shows the
surrounding wavelength region, including the two continuum windows
that were used to normalize the observed spectrum, while on the
{\it right panel}, a zoom on the Pb line is presented. 
The best fit is found for $\log \epsilon \mathrm{(Pb)}=-0.55$ (thick line)}
         \label{fig1}
   \end{figure*}

The signal we were looking for is very weak: a depression of only
$\simeq$0.5 per cent expected at 4057.807 \AA, in the red wing of a
weak CH line located at 4057.718 \AA. After correcting each spectrum
for radial-velocity shifts, several methods for combining the 13
spectra were tested, including (i) a straight average of the best 10
spectra (those with no cosmic hits in that wavelength region), (ii)
averaging the spectra after clipping points further away than
2.5$\sigma$ from the median of the distribution for each pixel, and
(iii) averaging the 9 spectra closest to the median of the
distribution for each pixel. All methods yielded a very similar result
in the Pb region, and we display only one of them (average of the nine
spectra closest to the median) in Fig. ~\ref{fig1}, where the error
bars represent the photon noise for each pixel.
 
\section{Spectral synthesis and comparison with observations}

We used the same model atmosphere and spectrum synthesis code 
({\it turbospectrum}: Plez et
al. \cite{PSL93}, Alvarez \& Plez \cite{AP98}) as in Paper~I, achieving 
complete self-consistency
between the model and the spectra computations. Our spectrum synthesis
of the Pb 4057\AA\ region is displayed in Fig. \ref{fig1}, where it is
compared to the observations. In addition to the photon noise itself
(error bars in the observed points in Fig.\ref{fig1}), various sources
of uncertainties on the Pb abundance determination were 
examined\footnote{
In addition, we noted a small, but clearly missing absorption component, in
the synthesis around 4057.63\AA, bluewards of the CH
line in Fig \ref{fig1}. This feature is too far from the Pb line to
affect its abundance, but to check the nature of this unidentified
feature, we compared our \object{CS\,31082-001} spectrum with the
spectra of C-rich stars of similar temperature, gravity, and
metallicity, and found that these stars also exhibited absorption
missing in the synthesis around 4057.63\AA, with an amplitude clearly
linked to the C and N abundances. We therefore conclude that it is a 
CH or CN molecular line.}:

{\it (i)} Continuum placement: there are two clean continuum
windows close to the Pb line, in the 4057.90 -- 4058.0\,\AA\ and
4058.4 -- 4058.5\,\AA\ intervals, which are used to achieve the best
normalization of the observations to the synthetic spectra. None of
the two windows are perfectly clean, the first containing a very
faint $\ion{Mn}{I}$ line at 4057.949\AA\ which seems slightly
underestimated in the synthesis, while the second has a $\ion{Co}{I}$
line at 4058.599\AA, slightly overestimated in the synthesis. The
two extreme normalizations differ by 0.17\%, which leads to a maximum
uncertainty on the Pb abundance of 0.15\,dex.

{\it (ii)} The wavelengths precision of the Pb and the blending
CH line at 4057.718\AA\ also affects the Pb abundance determination,
but to a much smaller extent. We have tested that a reasonable maximum
shift of 0.005\AA\ (0.35km/s) affects the Pb abundance by at most
0.05\,dex.

{\it (iii)} The isotopic ratio of $^{206}$Pb, $^{207}$Pb and
$^{208}$Pb can impact the Pb line shape, slightly changing its
effective central wavelength and hence affecting the abundance
determination. Following Van Eck et al. (\cite{vaneck2003}), we
considered five Pb components, one for each of the even isotopes 208
and 206 and the three hyperfine components for $^{207}$Pb. All
wavelengths and oscillator strengths were adopted from Van Eck et
al. (\cite{vaneck2003}). However, the two extreme cases of isotope
ratios that we have considered (see next section) did not produce any
noticeable difference in the spectrum synthesis and hence the derived
Pb abundances.

Considering the best fit spectrum displayed in Fig.~\ref{fig1}, and
the various sources of uncertainty outlined above, the Pb abundance in
\object{CS\,31082-001} is constrained to be $\log \epsilon\mathrm{(Pb)}
=-0.55\pm 0.15$\,dex, 0.35 dex below our upper limit in paper~I.

In the next section we discuss this result with respect to former
attempts to measure Pb abundance in other very old stars, and with
respect to the amount of lead expected from the decay of the actinides
Th and U. We do not attempt to explain our result by theoretical arguments, 
considering that it is the observational result that justify this Letter. 
A more complete discussion
of the impact of this new measurement on nuclear astrophysics will be
included in a forthcoming paper
also dealing with the analysis of our newly
completed HST/STIS observations of \object{CS\,31082-001}.

\section{Discussion} 

\subsection{Comparison with former observations of similar stars}

Other r-process enhanced stars have been
searched for Pb, leading only to upper limits or very uncertain
detections, whether using the very weak $\lambda$4058\AA\ line or the
intrinsically stronger (but observable only from space) UV line
$\lambda$2833\AA.
We report in Table~\ref{otherstars} the $\log$(Pb/Th) ratios
 for \object{CS~31082-001} and the two other stars with secure upper
limits: \object{CS~22892-052} (Th from Sneden et
al. \cite{SCL03}, Pb from Hill et al. \cite{HPC02}) and
\object{BD~+17\degr\,3248} (Cowan et al. \cite{CS02}). 
For completeness, we note that Sneden et al. (\cite{SCB98}) detected Pb
from the UV line in \object{HD~115444}, but the line was affected by a
spike and the authors themselves regarded the Pb abundance
in this star as very uncertain. 
Th is taken as reference element for the Pb abundance, because
of the direct connection between these elements. In the 3 stars, which
are as old or older than globular clusters, about half of the initial
Th content has decayed into $^{208}$Pb, and half has survived.  
The Pb/Th ratio detected in \object{CS~31082-001} clearly stands out as an
{\it extremely low value} compared to any upper limit so far placed on an
r-process enriched star.

\subsection{Comparison with the amount expected from the decay of $^{232}$Th,$^{235}$U and 
$^{238}$U}   
 Clear channels of production of lead are the decay of $^{232}$Th
into $^{208}$Pb , of $^{238}$U into $^{206}$Pb, and of $^{235}$U into
$^{207}$Pb. The amount of $^{208}$Pb is fixed by the observation of
$^{232}$Th now, and the knowledge of the decayed fraction after the
matter has been isolated in the atmosphere of the star, at a known
rate.

\begin{table}
\caption{Pb in \object{CS\,314082-001} and two other very metal-poor stars.}    
\label{otherstars}
\centering
\begin{tabular}{|l|l|l|l|}
\hline
Star &CS\,31082-001 &CS\,22892-052 &BD\,+$17\degr$\,3248\\
\hline 
[Fe/H] & $-$2.9 & $-$3.1 & $-$2.0  \\
$\log$Pb/Th &$0.43\pm0.16$ & $\leq 1.57$& $\leq 1.4$ \\
\hline
\end{tabular}
\end{table} 

The epoch of the nucleosynthesis of the photospheric matter of
CS\,31082-001, cannot be more than 13.7 $\pm$ 0.2 Gyr ago (Big Bang
epoch according to WMAP results, Spergel et al. \cite{SV03}), and
should be at least as much as the age of globular clusters (13.2 $\pm$
1.5 Gyr according to Chaboyer \cite{CHA01}), taking into
consideration the very low metallicity, $\mathrm{[Fe/H]}= -2.9$ 
of the star. The median is 13.5, also the age of first stellar 
formation according to Kogut et al. (\cite{KS03}).

 Adopting $t=13.5 \pm 1.5$~Gyr for the age of the actinides in CS
31082-001, we easily derive both the original content in $^{232}$Th
and $^{238}$U, and the fraction of them transformed into $^{208}$Pb
and $^{206}$Pb. For example:
$$\epsilon(^{206}\mathrm{Pb}) = \epsilon\mathrm{(^{238}U)_{now}}\times
(2^{(t/\tau)}-1)$$ with $\tau=4.47$ Gyr, the half-life of $^{238}$U. A
similar formula holds for $^{232}$Th and $^{208}$Pb with $\tau
=14.05$.
  
But $^{235}$U cannot be treated the same way, as there is not enough
$^{235}$U left to have an observed value. We must then rely on
theoretical works, usually done for reproducing the solar system
isotopic abundances, but not necessarily adequate for
\object{CS\,31082-001} which has a clear overabundance of the
actinides with respect to the lighter {\it r}-elements, compared to
the solar system.  However we can hope that in the restricted mass
range under consideration , $^{232}$Th to $^{238}$U, neutron exposures
producing the right ratio $^{238}$U/$^{232}$Th may also produce the
right ratio $^{235}$U/$^{238}$U.  In Tables 1 and 2 of Goriely \&
Arnould (\cite{GA01}), several neutron exposures are considered with a
wide set of mass models. Forgetting the solar system, we keep the
exposures giving the right $^{238}$U/$^{232}$Th production ratio for
CS 31082-001, compatible with an age 13.5 $\pm$ 1.5 Gyr. With this
constraint, the production ratio $R$ of $^{235}$U/$^{238}$U lies
between 0.67 and 0.87.  Taking the median 0.77 seems a reasonable
estimate. The full amount of produced $^{235}$U is converted into
$^{207}$Pb in \object{CS31082-001}, because of the fast decay of this
isotope.  Table~\ref{isotopes} summarizes our findings. Interestingly,
the case $R=1.0$ gives an amount of total lead equal to the observed
one, leaving no other channel for the production of {\it r}-lead. We
have not included the $^{234}$U decay to $^{206}$Pb channel here, 
because of the lack of published estimates of the $^{234}$U/$^{238}$U 
production ratio.
This channel (which could 
be as high as the $^{235}$U contribution) can only
increase the contribution of the actinide-path to the total production
of Pb, thereby reducing even further any other production 
channel.

A warning is appropriate here: our analysis is based on the LTE
approximation, which must be questioned, especially in the blue and
the UV in very metal-poor giants, where the continuum is in a large 
part due to Rayleigh scattering. We examine this in the next subsection.

\begin{table}
\centering
\caption{Three abundance patterns examined for the lead feature.   
The abundances of $^{207}$Pb are computed for 3 assumed values of the 
production ratio R=$^{235}\mathrm{U}/^{238}\mathrm{U}$, the first 
considered as the most probable, and the other two chosen
 30 per cent below and above. All
abundances are with respect to $10^{12}$ hydrogen atoms.
}
\label{isotopes}
\begin{tabular}{|l|l|l|l|}
\hline\hline 
$^{238}$U$_{\mathrm{now}}$ &  \multicolumn{3}{c|}{0.0120$\pm$ 0.0035} \\
$^{206}$Pb &  \multicolumn{3}{c|}{0.086$\pm$0.036}\\
$^{232}$Th$_{\mathrm{now}}$ &\multicolumn{3}{c|}{ 0.105$\pm$.013 }\\
$^{208}$Pb &  \multicolumn{3}{c|}{0.099$\pm$0.035}\\
\hline
    R      &    0.77  & 0.59  &  1.00  \\
$^{207}$Pb &   0.075 $\pm$.03  &  0.058$\pm$.02 &  0.098$\pm$.04\\
tot. Pb &0.26$\pm$0.10  & 0.243$\pm$0.09    & 0.283$\pm$.11 \\
$\log(\epsilon$(Pb)) & $-$0.59$\pm$.2 & $-$0.61$\pm$.2   & $-$0.55$\pm$.2 \\
\hline
\end{tabular}
\end{table}    

\subsection{NLTE versus LTE}
It would be very useful to have a NLTE analysis of Pb, as the lower
level of the measured line is very deep, part of the ground-level
term, and is of Pb I when most of lead is in the Pb II stage. In
metal-poor stars these deep levels tend to be over-ionized, but not
always. If the deep levels are indeed over-ionized, the LTE assumption
predicts too large a population of the lower level, and an
underestimated abundance. In an attempt to estimate the size of
possible NLTE effects, we checked the J$_\nu$/B$_\nu$ ratio at the
position of the Pb~I line, as well as at 4076\AA\ and 2035\AA,
corresponding to the ionization limit of the upper and lower levels of
the transition, respectively. In these layers, the ratio
J$_\nu$/B$_\nu$ of the mean intensity to the local Planck function is
larger than one, but remains on the order of two.  It is unlikely that
the NLTE correction is larger than this factor of two, so the
production of Pb, outside the decay of $^{232}$Th, $^{235}$U, and
$^{238}$U is bound to be less or equal to the actinide production.

\subsection{Digression about the rapid neutron capture lead in the solar system}
Our result clearly concerns a particular class of objects: very
metal-poor stars born in the early days of the Galaxy, and strongly
enhanced in {\it r}-process elements.  Is the observed low ratio Pb/Th
particular to this class of objects, or is it a more general property
of the {\it r}-process in the mass range 206-238? There is now a
general agreement that the fraction of Pb produced by the {\it r}-process
in the solar system it practically unknown, after the discovery that
zero-metal stars can produce a lot of {\it s}-lead (Goriely \& Siess
\cite{GS01}, Van Eck et al. \cite{vaneck2003}). This has modified the
estimates of the amount of {\it s}-lead produced in the Galaxy before
the birth of the Sun (Gallino et al \cite{GAB98}), and of the {\it
r}-lead, obtained by subtracting the {\it s}-lead from the total lead.
If the {\it r}-lead of the solar system were mainly produced by the
decay of the actinides, as for CS 31082-001, it is easy to verify that
the {\it r}-lead in the solar system would be of the order of only 1
to 3 \% of the total lead.

\section{Conclusions}
In one of the very metal-poor stars showing a large enhancement of {\it
r}-process elements we have now a true determination of the lead
abundance, instead of upper limits, only.  This abundance is very
low,$-0.55 \pm 0.15$\ dex in \object{CS\,31082-001}, about one dex
below the former upper limits in \object{CS\,22892-052} and
\object{BD\,+17\degr\,3248}.  Also, our result shows that, in the
purely $r$-process enriched photosphere of \object{CS\,31082-001},
most of lead results from the decay of $^{232}$Th , $^{235}$U, and
$^{238}$U.  This places a limit on the amount of $^{235}$U which has
contributed to the production of $^{207}$Pb, as well as on that
of $^{234}$U. A non-LTE analysis of the
spectrum is highly desirable, but hampered so far by the lack of
photoionization cross-sections for Pb I.  

\acknowledgements We are
indebted to Prof. R. Gallino for informations on the production of
lead by the {\it s}-process.  T.C.B. acknowledges partial funding from
NSF grants AST 00-98508 and 00-98549, and PHY 02-16783.BN acknowledges 
support from the Carlsberg Foundation and the Nordic Academy for 
Advanced Studies.


\begin{thebibliography}{}
\bibitem[1998]{AP98}
Alvarez, R., \& Plez, B. 1998, \aap \ 330, 1109
\bibitem[2000]{bal00} 
Ballester, P., Modigliani, A., Boitquin, O., et al. 2000, 
The Messenger, 101, 31
\bibitem[2001]{CHA01}
Chaboyer, B. 2001, in {\it Astrophysical ages and 
time scales} Ed. T. von Hippel et al., 
 ASP conf.series vol. 245, p.162
\bibitem[2001]{CHB01} 
Cayrel, R., Hill, V., Beers, T.C. et al. 2001, Nature, 409, 691
\bibitem[2002]{CS02}
Cowan, J.J., Sneden C., Burles, S. et al. 2002, \apj \ 572, 861
\bibitem[1998]{GAB98}
Gallino, R., Arlandini, C., Busso, M., et al.
1998, \apj \ 497, 388
\bibitem[2001]{GA01}
Goriely, S., \& Arnould, M. 2001, \aap \ 379, 1113
\bibitem[1999]{GC99}
Goriely, S., \& Clerbaux B. 1999, \aap \ 346, 798
\bibitem[2001]{GS01}
Goriely, S., Siess, L. 2001, \aap\ 378, L25
\bibitem[2002]{HPC02}
Hill, V., Plez,B., Cayrel, R., et al.
2002, \aap \ 387, 560 (paper I)
\bibitem[2003]{KS03}
Kogut, A., Spergel, D.N., Barnes,C. 2003, \apjs \ 148, 161
\bibitem[2002a]{NIJL02}
Nilsson, H., Ivarsson, S., Johansson, S., \& Lundberg, H., 2002, \aap\ 381, 1090
\bibitem[2002b]{NZL02}
Nilsson, H., Zhang, Z.G., Lundberg, H., et al. 2002, \aap\ 382, 368 
\bibitem[1993]{PSL93}
Plez, B., Smith, V.V., Lambert, D.L. 1993, \apj \ 418, 812
\
\bibitem[1998]{SCB98}
Sneden, C., Cowan, J.J., Burris, D.L. et al. 1998, \apj\ 496, 235
\bibitem[2003]{SCL03}
Sneden, C., Cowan, J.J., Lawler, J.E. et al. 2003, \apj\ 591, 936
\bibitem[2003]{SV03}
Spergel, D.N., Verde, L., Peiris, H.V. et al. 2003, \apjs \ 148, 175
\bibitem[2003]{vaneck2003}
Van Eck, S., Goriely, S., Jorissen, A., Plez, B., 2003, \aap \ 404, 291
\bibitem[2002]{WII02}
 Wanajo, S., Itoh, N., Ishimaru, Y., Nozawa, S., Beers, T.~C., 2002,
 \apj\  577, 853
\end{thebibliography}
\end{document}